\documentclass[aps,prb,preprint,graphicx,amsbsy,amssymb]{revtex4}

\usepackage{graphicx}
\usepackage{graphics}
\usepackage{epsfig}

\begin{document}
\title{ Spinon-Holon binding in $t-J$ model }

\author{Tai-Kai Ng}
\address{Department of Physics, Hong Kong University of Science and
Technology, Clear Water Bay Road, Hong Kong}

\begin{abstract}

  Using a phenomenological model, we discuss the consequences of spinon-holon binding in the $U(1)$ slave-boson
 approach to $t-J$ model. Within a small $x$ ($x=$ hole concentration) expansion, we show that spinon-holon binding
 produces a pseudo-gap normal state with a segmented Fermi surface and the superconducting state is formed by
 opening an "additional" d-wave gap on the segmented Fermi surface. The d-wave gap merge with the pseudo-gap
 smoothly as temperature $T\rightarrow0$. The quasi-particles in the superconducting state are coupled to external
 electromagnetic field with a coupling constant of order $x^{\gamma}$ where $0\leq\gamma\leq1/2$, depending on the
 strength of the effective spinon-holon binding potential.

  \end{abstract}

\pacs{74.20.Rp, 74.45.+c, 74.50.+r}

\maketitle

\narrowtext

   The $U(1)$ slave-boson mean-field theory (SBMFT) of the $t-J$ model has been used by many authors as a starting
 point for the theory of high-$T_c$ superconductors\cite{t1,t2,t3,t4}. With suitable refinements the theory can
 explain a lot of the qualitative features of the cuprates\cite{t2,t3,t4,t5}. However the theory does not produce a
 satisfactory description of the low energy quasi-particle properties in the under-doped regime. It predicts a very
 strong renormalization of quasi-particle charge in the superconducting state\cite{t5,t6} which is not observed
 experimentally\cite{e1,e2}. It has been suggested\cite{t3,t7} that the failure of SBMFT in describing
 quasi-particles is due to the lack of consideration of confinement between low energy spinons and holons coming
 from strong gauge field fluctuations. This scenario has been studied in the $SU(2)$ formulation of $t-J$ model
 where it was found that spinon-holon binding leads to formation of half-pocket (segmented) Fermi surfaces in the
 normal state\cite{t3,t7}, and a rather normal d-wave superconductor state. In this paper we study the effect of
 spinon-holon binding in the $U(1)$ slave-boson formulation of $t-J$ model, assuming that an effective spinon-holon
 interaction which is constant at distance range $d<l\sim \pi^{-1}x^{-1/2}$ exists. The effective interaction
 strength $U_o(x)$ is a phenomenological $x$-dependent parameter in our theory. The purpose of our paper is to study
 how the quasi-particle properties in SBMFT are modified in the presence of this phenomenological interaction in the
 small $x$ limit (under-doped regime), and to compare with experiments\cite{pe,tu}.

  We consider a model Hamiltonian on a two dimensional square lattice, $H=H_{MF}^f+H_{MF}^b+H_{c}$, where
 \begin{equation}
 \label{hmfs}
 H_{MF}^f=\sum_{\vec{k}\sigma}\xi_{\vec{k}}f^+_{\vec{k}\sigma}
 f_{\vec{k}\sigma}+\sum_{\vec{k}}\left[\bar{\Delta}^*(\vec{k})(f_{\vec{k}\uparrow}
 f_{-\vec{k}\downarrow}-f_{\vec{k}\downarrow}f_{-\vec{k}\uparrow})+H.C.\right],
 \end{equation}
 is the fermion (spinon) mean-field Hamiltonian in SBMFT.
 $\xi_{\vec{k}}=-(t\bar{b}^2+{3J\over8}\bar{\chi})\gamma(\vec{k})-\mu_f$, $\bar{\chi}=<\sum_{\sigma}f^+_{i\sigma}
 f_{j\sigma}>$ where $j=i+\hat{\mu} (\mu=x,y)$, and $\gamma(\vec{k})=2(\cos(k_x)+\cos(k_y))$.
 $\Delta(\vec{k})={3J\over4}\bar{\Delta}(\cos(k_x)-\cos(k_y))$ is a spinon pairing field where
 $\bar{\Delta}=<f_{i\uparrow}f_{j\downarrow}-f_{i\downarrow}f_{j\uparrow}>$. The mean-field dispersion for the
 spinon is $E_f(\vec{k})=\pm\sqrt{\xi_{\vec{k}}^2+|\Delta(\vec{k})|^2}$\cite{t1,t5}.

  The boson (holon) mean-field Hamiltonian is
 \begin{equation}
 \label{hmfh}
 H_{MF}^h=\sum_{\vec{q}}\epsilon(\vec{q})b^+_{\vec{q}}b_{\vec{q}}
 +{U_h\over2}\sum_{\vec{q}}\Delta_h(b^+_{\vec{q}}+b_{-\vec{q}})(b_{\vec{q}}+b^+_{-\vec{q}}),
 \end{equation}
 where we have introduced a a short-ranged hole-hole repulsion term $U_h\sim t$ which is treated by usual
 Bogoliubov approximation\cite{mahan}.
 \[
  \Delta_h=\bar{x}^2+{1\over V}\sum_{\vec{q}}<b_{\vec{q}}b_{-\vec{q}}>,  \]
 where $\bar{x}$ is the hole Bose-condensation amplitude. Note that $\bar{x}<x$ in the
 presence of hole-hole and holon-spinon interactions. $\epsilon(\vec{q})=-t\bar{\chi}\gamma(\vec{q})+\mu_b$. The
 existence of Bose-condensation $\bar{x}>0$ implies $\mu_b=t\bar{\chi}\gamma(0,0)$. The mean-field dispersion for
 the holon excitation is $E_h(\vec{q})=\sqrt{\epsilon(\vec{q})^2+2\epsilon(\vec{q})U_h\Delta_h}$\cite{t5}.

 We assume an effective spinon-holon interaction of form
 \begin{equation}
 \label{hbf}
 H_{c}=U_o(x)\sum_{(|\vec{k}-\vec{k}|_F,|\vec{q}|,|\vec{q}'|<\Lambda)\sigma}(f^+_{\vec{k}+\vec{q}\sigma}b_{\vec{q}})
 (b^+_{\vec{q}'}f_{\vec{k}+\vec{q}'\sigma}),
 \end{equation}
 where $U_o(x)\sim t/ x^{-\eta}$ is a phenomenological spinon-holon interaction and $\Lambda^{-1}\sim\pi\sqrt{x}$.
 We assume that the binding potential is effective only at a small range of momentum $\leq\Lambda$ around
 $\vec{q}=0$ and around the spinon Fermi surface. Gaussian fluctuations above SBMFT produces an effective spinon-holon
 interaction with $\eta=0$\cite{t5}. The stronger effect of confinement is mimicked by a potential with $\eta>0$. We
 shall treat $U_o$ as a phenomenological parameter and shall examine its effect for various values of $\eta$ in the
 following.

   The electron Green's function is computed in a Generalized self-consistent Born Approximation that involves
 self-consistent evaluation of the electron and boson Green's functions,
 \begin{eqnarray}
 \label{green}
 \bar{G}_c(k)=\pmatrix{ G_c(k) & F_c(k) \cr F^*_c(k) &
 -G_c(-k)},\,\
 \bar{G}_b(q)=\pmatrix{ G_b(q) & F_b(q) \cr F^*_b(q) & G_b(-q)}
 \end{eqnarray}
 where $k=(\vec{k},i\omega)$. The self consistent equation is
 \begin{mathletters}
 \label{gscc}
 \begin{equation}
 \label{gc}
 \bar{G}_c(k)=\bar{G}_c^{(0)}(k)+[U_o\bar{G}_c(k)],
 \end{equation}
 where
 \begin{eqnarray}
 \label{g0}
 \bar{G}_c^{(0)}(k) & = & \pmatrix{g_c(k) & f_c(k)
 \cr f_c^*(k) & g_c(-k)},
 \end{eqnarray}
 and
\begin{eqnarray}
\label{gf0}
 g_c(k) & = & xZ_g(T)g_f(k)+{1\over V\beta}\sum_{q}g_f(k+q)G_b(q), \\  \nonumber
 f_c(k) & = & xZ_f(T)f_f(k)-{1\over V\beta}\sum_{q}f_f(k+q)F_b(q)
 \end{eqnarray}
 where $g_f(k)={i\omega+\xi_{\vec{k}}\over i\omega^2-E_f(\vec{k})^2}$ and
 $f_f(k)={-\Delta_{\vec{k}}\over i\omega^2-E_f(\vec{k})^2}$ are the mean-field normal and
 anomalous Green's functions of the spinons, respectively. $\sqrt{xZ_{g(f)}(T)}$ are effective Bose-condensation
 amplitudes in the normal and anomalous Green's functions and
 \begin{eqnarray}
 \label{ug}
 [U_oG_c(k)] & = & \bar{\theta}(\vec{k}){U_o\over V\beta}\sum_{|\vec{q}|<\Lambda}g_f(k+q)G_b(q)
 \\  \nonumber
 [U_oF_c(k)] & = & \bar{\theta}(\vec{k}){U_o\over V\beta}\sum_{|\vec{q}|<\Lambda}f_f(k+q)F_b(q)
 \end{eqnarray}
 \end{mathletters}
 where $\bar{\theta}(\vec{k})=\theta(\Lambda-|\vec{k}-\vec{k}_F|)$. The boson Green's functions are given by
 \begin{mathletters}
 \label{gscb}
 \begin{eqnarray}
 \bar{G}_b^{-1}(q) & = & \pmatrix{i\Omega-\epsilon_{\vec{q}}-\Sigma_b(q) & \Delta_b(q)
 \cr \Delta^*_b(q) & -i\Omega-\epsilon_{-\vec{q}}-\Sigma_b(-q)},
 \end{eqnarray}
 where
 \begin{eqnarray}
 \label{selfb}
 \Sigma_b(q) & = & U_h\Delta_h+{U_o^2\over V\beta}\sum_{|\vec{k}|-k_f<\Lambda}G_c(k)g_f(k+q),
 \\ \nonumber
 \Delta_b(q) & = & U_h\Delta_h-{U_o^2\over
 V\beta}\sum_{|\vec{k}|-k_f<\Lambda}F^*_c(k)f_f(k+q).
 \end{eqnarray}
 \end{mathletters}
 The system is in a superconducting state if both the {\em electron} normal and anomalous Green's functions are
 nonzero, and is in the normal state if only the normal Green's function is nonzero. Self-consistent determination
 of the Green's functions is not carried out in the original $SU(2)$ theory\cite{t3,t7}. We find that self-consistent
 determination of electron and boson Green's function is important in determining the quasi-particle properties
 whereas self-consistency for the spinon Green's function does not affect the qualitative properties of the system.
 The spinon Green's function is not evaluated self-consistently for simplicity.

   Equations (4) to (10) can be solved in a small-$x$ expansion when $\eta<1/2$. First we consider zero
 temperature. In this case $xZ_g(0)=xZ_f(0)=\bar{x}\sim x$ and the electron Green's functions can be written as
 \begin{eqnarray}
 \label{egreen}
 G_c(k) & \sim & G_{inc}(k)+{\bar{x}(\omega+\bar{\xi}_{\vec{k}})\over\omega^2-\bar{E}^2_{\vec{k}}}
 +O(x^{{\gamma\over2}})   \\  \nonumber
 F_c(k) & \sim & -{\bar{x}\Delta_{\vec{k}}\over\omega^2-\bar{E}^2_{\vec{k}}}+O(x^{{\gamma\over2}})
 \end{eqnarray}
 where $\gamma=1-\eta$ for $\eta<1/4$ and $\gamma=3({1\over2}-\eta)$ for $\eta>1/4$. $\bar{\xi}_{\vec{k}}=
 \xi_{\vec{k}}+U_o\bar{x}$ and $\bar{E}^2_{\vec{k}}=\bar{\xi}_{\vec{k}}^2+\Delta_{\vec{k}}^2$. $G_{inc}$ is
 a smooth background coming from convolution of the spinon and holon Green's functions. The quasi-particle behavior
 is determined by the terms with weight $\sim\bar{x}$. For $|\omega|\sim \bar{E}_{\vec{k}}\lesssim U_o\bar{x}$, the
 quasi-particle term is of order $O(1)$ and is larger than the $O(x^{{\gamma\over2}})$ term for $\eta<1/2$. This is
 also the energy range where the spinon-holon bound states are stable. The bound states become unstable and decay
 into separate spinons and holons at $\bar{E}_{\vec{k}}\geq U_o\bar{x}$.

   At temperatures $T>T_c\sim T_{BE}$ where Bose-condensation vanishes, the anomalous Green's function is zero. For
 $T$ just above $T_{BE}$ we may consider the bosons to be "almost Bose-condensed"\cite{t3} and $Z_g(T\geq
 T_{BE})\sim 1$ and $Z_f(T\geq T_{BE})\sim 0$. We obtain with this approximation
 \begin{equation}
 \label{egreen2}
 G_c(k)\sim G_{inc}(k)+{x(\bar{\omega}+\bar{\xi'}_{\vec{k}})\over\bar{\omega}^2-\bar{E'}^2_{\vec{k}}}
 +O(x^{{\gamma\over2}})
 \end{equation}
 where $\bar{\xi'}_{\vec{k}}=\xi_{\vec{k}}+U_ox/2$, $\bar{\omega}=\omega-U_ox/2$ and
 $\bar{E'}^2_{\vec{k}}=\bar{\xi'}_{\vec{k}}^2+\Delta_{\vec{k}}^2$. Notice that for each momentum $\vec{k}$ there
 exists two branches of quasi-particles with energies $\pm \bar{E}$, both in the superconducting and the normal states.

  The transition from superconducting to normal state as temperature raises from $0$ to $T_c$ can also be studied in
  the "almost Bose-condensed" approximation where we approximate $Z_g(T)\sim 1$ and $1>Z_f(T)>0$. With this we
  obtain after some algebra,

 \begin{equation}
 \label{egreen3}
 G_c(k)\sim G_{inc}(k)+{x\over4}\sum_{i=1,2}\left({u_i(k)\over\omega-
 E_i(\vec{k})}+{v_i(\vec{k})\over\omega+E_i(\vec{k})}\right)+O(x^{{\gamma\over2}})
 \end{equation}
 where $E_{1(2)}^2(\vec{k})=Z_f\Delta_{\vec{k}}^2+(E^c_{\vec{k}}+(-){U_ox\over2})^2$,
 $E^c_{\vec{k}}=\sqrt{\bar{\xi'}^{2}_{\vec{k}}+(1-Z_f)\Delta^2_{\vec{k}}}$, and
 $u(v)_1(\vec{k})=(1+{\bar{\xi'}_{\vec{k}}\over E^c_{\vec{k}}})\left(1+(-){\bar{\xi'}_{\vec{k}}+U_ox/2\over
 E_1(\vec{k})})+(-)(1-Z_f){\Delta^2_{\vec{k}}\over E^c_{\vec{k}}E_1(\vec{k})}\right)$,
 $u(v)_2(\vec{k})=(1-{\bar{\xi'}_{\vec{k}}\over E^c_{\vec{k}}})\left(1+(-){\bar{\xi'}_{\vec{k}}+U_ox/2\over
 E_2(\vec{k})})-(+)(1-Z_f){\Delta^2_{\vec{k}}\over E^c_{\vec{k}}E_2(\vec{k})}\right)$. A corresponding expression
 also exists for $F_c(k)$.

   We observe that four branches of quasi-particles exist at intermediate temperature $0<T<T_c$. However $u_2=v_2=0$
 at $T=0$ and $v_1=u_2=0$ at $T\geq T_c$, showing that two of the four branches of quasi-particles vanish at these
 temperatures. These results together suggest a rather unconventional picture for the low temperature behavior of
 high-$T_c$ cuprates in our theory. The normal state is a pseudogap state where the tunneling density of states
 (DOS) has a d-wave like gap in the spectrum. Notice that unlike the DDW state\cite{ddw}, the present pseudo-gap
 state does not break translational symmetry. The normal density of states computed by Eq.\ (\ref{egreen2}) for
 three different dopings $x=0,0.05,0.1$ with $J=1, t=3$ and $U_h=U_o=4 (\eta=0)$ and energy resolution
 $\Delta\omega\sim0.15J$ is shown in figure 1. The cutoff in momentum space is implemented by introducing a cutoff
 factor $e^{-q^2/\Lambda^2}$ in the numerical integrations. The shifting of DOS from low to high energy into
 quasi-particle spectral weight as $x$ increases with a d-wave gap structure is clear from the figure. These global
 features are in agreement with tunneling measurements\cite{tu}.The DOS in the superconducting
 state is similar to the normal state except the sharper d-wave gap structures and that the superconducting DOS
 minimum is pinned at zero energy whereas it is pinned at an energy $E\sim U_ox/2$ in the normal state.
 \begin{figure}
 \includegraphics[width=6.0cm, angle=-90]{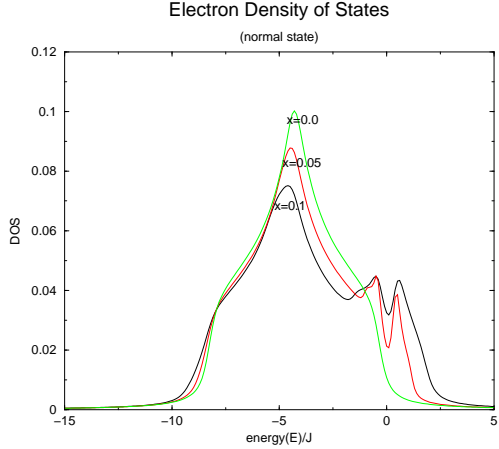}
 \caption{\label{Fig.1} Tunneling density of states for $x=0,0.05,0.1$.}
 \end{figure}
 Correspondingly, the position of the Fermi surface in the normal state defined by $\bar{E'}_{\vec{k}}=-U_ox/2$ is
 down shifted from the nodal point $\bar{E'}_{\vec{k}}=0$ resulting in formation of segmented Fermi surface below
 the nodal points as in $SU(2)$ theory\cite{t3,t7}. We show in figure two the electron normal state occupation
 number computed at $x=0.1$. The electron occupation number exhibit "pocket" structures although the discontinuity
 across the Fermi surface seems to occur only on the inner side of the pocket (fermi arc), in agreement with
 experiments\cite{pe}.
 \begin{figure}
 \includegraphics[width=8.0cm, height=5.0cm]{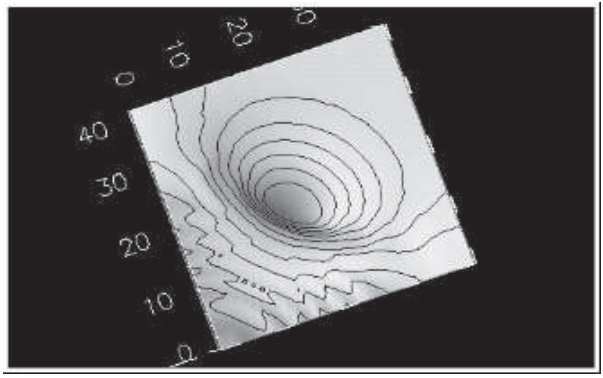}
 \caption{\label{Fig.2} Electron occupation in the normal state for $x=0.1$ at the momentum region
 $0\leq k_x,k_y \leq \pi$. The lines are equally spaced constant contours with $n_{max}\sim0.52$ and $n_{min}\sim0.31$}
 \end{figure}

   The superconducting transition at $T_c$ is driven by opening an "additional" {\em superconducting} gap
 $(\sim Z_f^{1/2}\Delta_{\vec{k}})$ at the Fermi arc. The opening of the superconducting gap leads to
 formation of four branches of quasi-particles at intermediate temperature below $T_c$. The superconducting order
 parameter merges with the pseudo-gap smoothly as temperature $T\rightarrow0$ where two of the four branches of
 quasi-particles vanish. In figure (3) we show the electron density of states computed in our theory at $x=0.1$
 for three different values of $Z_f=0.0$ (normal state), $Z_f=0.5$ ($T>0$ superconducting state) and $Z_f=1.0$
 ($T=0$ superconducting state). The gradual opening of gap at the Fermi arc ($Z_f=0.5$) and merging of
 superconducting gap with pseudo-gap at $T=0 (Z_f=1)$ is clear from the figure. The smooth merging of the
 superconducting gap with the pseudo-gap reflects the common origin of the two gaps in our theory and would be
 absent in other theories where the two gaps are of different origins.
 \begin{figure}
 \includegraphics[width=6.0cm, angle=-90]{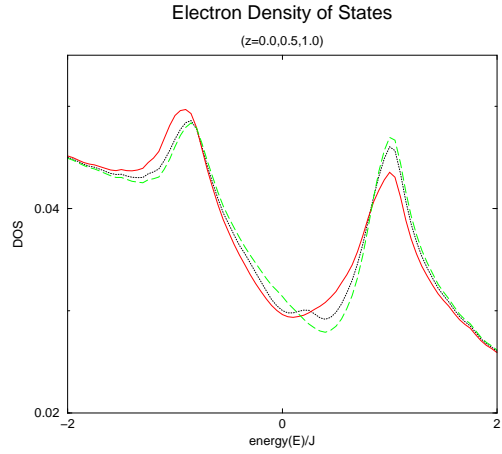}
 \caption{\label{Fig.3} Electron density of states at $x=0$ for three different values of $z=0.0,0.5,0.1$}
 \end{figure}

   We now provide some technical details of our self-consistent calculation. In the small $\vec{q},\Omega$ limit
 we may expand the holon Green's function self-energies to obtain
 \begin{eqnarray}
 \label{bgreen}
 G_b(q) & = & {\bar{\Omega}+\bar{\epsilon}_{-\vec{q}}\over(\bar{\Omega}-\bar{\epsilon}_{\vec{q}})
 (\bar{\Omega}+\bar{\epsilon}_{-\vec{q}})+|\Delta_b(0)|^2},  \\
 \nonumber
 F_b(q) & = & {-\Delta^*_b(0)\over(\bar{\Omega}-\bar{\epsilon}_{\vec{q}})
 (\bar{\Omega}+\bar{\epsilon}_{-\vec{q}})+|\Delta_b(0)|^2},
 \end{eqnarray}
 where $\bar{\Omega}=Z^{-1}_b\omega$, $\Delta_b(0)=\Delta_b(\vec{q}=0,\Omega=0)$,
 $Z^{-1}_b=1-\left({\partial\Sigma_b(\vec{q}=0,\omega)\over\partial\omega}\right)_{\omega=0}$ and
 $\bar {\epsilon}_{\vec{q}}=\epsilon_{\vec{q}}+\left({\partial\Sigma_b(\vec{q},\omega=0)\over\partial(\vec{q})^2}
 \right)_{\vec{q}=0}\vec{q}^2\sim t_{eff}\vec{q}^2-\mu_h$. A consistent evaluation of the self-energies gives
 ${\partial\Sigma_b\over\partial\omega}\sim(U_0^2/J)\bar{x}$, ${\partial\Sigma_b\over\partial(\vec{q})^2}
 \sim U_0$ and $\Delta_b(0)\sim U_h\Delta_h-(U_o^2/J)\bar{x}^{{3\over2}}$. Therefore, $t_{eff}\sim t+U_o$
 and $Z_b\sim 1$ for $\eta<1/2$ at small $x$. At zero temperature the boson occupation number is a sum of two terms,
 $x=\bar{x}+x_{nbc}$, where $x_{nbc}\sim \sqrt{\Delta_b(0)\bar{x}\over t_{eff}}$ is the density of uncondensed
 bosons arising from holon-holon and spinon-holon interactions. With \ (\ref{bgreen}) we obtain a
 self-consistent equation for $\Delta_b(0)$ of form
 \[
 \Delta_b(0)\sim a\bar{x}^{{1\over2}(1+\eta)}\sqrt{\Delta_b(0)}+U_h\bar{x}-b\bar{x}^{{3\over2}-\eta},
 \]
 where $a\sim U_h,b\sim t^2/J$ are numerical factors. It is easy to see that $\Delta_b(0)\sim \bar{x}$ for
 $\eta<1/4$ when holon-holon repulsion dominates and $\Delta_b(0)\sim \bar{x}^{{3\over2}-2\eta}$ for $\eta>1/4$
 where spinon-holon interaction dominates. Correspondingly, $x_{nbc}\sim\bar{x}^{1+{\eta\over2}}$ for $\eta<1/4$
 and $x_{ncb}\sim\bar{x}^{1+{1\over2}({1\over2}-\eta)}$ for $\eta>1/4$. In particular, $x_{ncb}>\bar{x}$ at
 $\eta>1/2$ where Bose-condensation vanishes and our small $x$ expansion which assumes $\bar{x}\sim x$ breaks
 down. In this case, a new state that cannot be described by SBMFT as starting point is formed.

  A problem associated with the SBMFT approach to high-$T_c$ superconductors is that the quasi-particle
 charge $q_{eff}$ inferred from the temperature dependent London penetration depth is of order $O(1)$
 experimentally\cite{e1}, but is of order $x$ is SBMFT\cite{t5,t6}. A recent experiment also indicates that the
 simple quasi-particle picture for temperature dependent London penetration depth\cite{t8} may be violated at the
 extremely low-doping regime\cite{e2}. In the following we study how quasi-particles in our theory couple to external
 electromagnetic field. The electromagnetic field $\vec{A}$ couples to electrons through the $t$-term in the $t-J$
 model. The paramagnetic coupling to electromagnetic field is in linear response,
 \[
 \delta{H}_P=-2it\sum_{\vec{k}.\vec{q}}\sin(k_{\mu})
 \left(\bar{x}\sum_{\sigma}f^+_{\vec{k}+\vec{q}\sigma}f_{\vec{k}\sigma}+\bar{\chi}b^+_{\vec{k}+\vec{q}}b_{\vec{k}}\right)
 A_{\mu}(\vec{q}),  \]
 where the first term represents coupling through spinons and the second term represents coupling through holons.
 The paramagnetic coupling of $\vec{A}$ to quasi-particles (= spinon-holon bound state in our theory) can be evaluated
 by examining the first order change in electron Green's functions by $\delta{H}_P$. After some lengthy algebra, we
 find that at zero temperature the nodal quasi-particles in the superconducting state are minimally coupled to
 external magnetic field as in usual d-wave superconductors with coupling constant $q_{eff}\sim
 (U_o/ J)(x_{ncb}/\bar{x}^{1/2})\sim x^{\gamma\over2}$. We note that $\gamma>0$ for $1/2>\eta>0$ and $q_{eff}$
 still vanishes as $x\rightarrow0$, although the vanishing rate is slower than predicted in pure SBMFT.
 The quasi-particle charge becomes of order $O(1)$ only when $\eta\rightarrow1/2$, where Bose-condensation vanishes.

  Summarizing, we have examined the effect of spinon-holon binding on the quasi-particle properties in the $U(1)$
  slave-boson approach to $t-J$ model. Within a small-$x$ expansion we find that spinon-holon binding produces an
  effective low energy theory with a segmented Fermi surface at normal state and a rather normal d-wave
  superconductor at $T=0$, in agreement with photo-emission experiments\cite{pe}. The theory predicts a rather
  non-trivial crossover behavior between $T=T_c$ to $T=0$ where four branches of quasi-particles exist in the
  intermediate state. This prediction is yet to be tested in photo-emission or tunnelling experiments.
  The problem of quasi-particle charge is not completely resolved in our theory. Our analysis suggests that
  quasi-particle charge is of order one only when the confining potential is strong enough such that
  Bose-condensation amplitude vanishes and our analysis breaks down. A fully self-consistent treatment
  in this regime is still missing and will be the subject of future investigations.

{\it Acknowledgements}
 The author thanks P.A. Lee for interesting him in this investigation and for many useful comments.
 This work is supported by HKRGC through Grant 602803.

 \end{document}